\documentclass[acmsmall,screen,nonacm]{acmart}

\usepackage[utf8]{inputenc}
\usepackage[T1]{fontenc}
\usepackage{amsmath,amsfonts}
\usepackage{textcomp}
\usepackage{hyperref} 
\graphicspath{{./img}}
\usepackage{enumitem}
\usepackage[frozencache]{minted}
\usepackage{minted}
\setminted[prolog]{frame=lines,breaklines,fontsize=\footnotesize}

\newif\ifcommentson
\commentsontrue

\newlist{enum*}{enumerate*}{1}
\setlist*[enum*,1]{%
  label=(\arabic*),
}

\begin{document}

\title{Unveiling the Mechanisms of DAI: A Logic-Based Approach to Stablecoin Analysis
}

\author{Francesco De Sclavis}
\orcid{0009-0004-1318-3878}
\author{Giuseppe Galano}
\orcid{0009-0008-3251-6606}
\author{Aldo Glielmo}
\orcid{0000-0002-4737-2878}
\author{Matteo Nardelli}
\orcid{0000-0002-9519-9387}


\begin{abstract}
Stablecoins are digital assets designed to maintain a stable value, typically pegged to traditional currencies. Despite their growing prominence, many stablecoins have struggled to consistently meet stability expectations, and their underlying mechanisms often remain opaque and challenging to analyze. 
This paper focuses on the DAI stablecoin, which combines crypto-collateralization and algorithmic mechanisms. We propose a formal logic-based framework for representing the policies and operations of DAI, implemented in Prolog and released as open-source software.
Our framework enables detailed analysis and simulation of DAI’s stability mechanisms, providing a foundation for understanding its robustness and identifying potential vulnerabilities.
\end{abstract}

\maketitle

\section{Introduction}

Stablecoins have emerged as a cornerstone of the cryptocurrency ecosystem, promising to mitigate the price volatility commonly associated with digital assets \cite{mita2019stablecoin}. By maintaining a stable value, typically pegged to a fiat currency like the US dollar, they aim to combine the reliability of traditional currencies with the benefits of blockchain technology. However, the extent to which stablecoins fulfill this promise is a subject of ongoing scrutiny. 
Indeed, stablecoins can deviate from their intended price pegs\footnote{This deviation can also be dramatic, because of certain events, as happened to USDC: \url{https://www.wsj.com/articles/crypto-investors-cash-out-2-billion-in-usd-coin-after-bank-collapse-1338a80f}.}, and this has prompted research into the factors determining their stability or instability (e.g., see~\cite{Potter24,duan2023instability}).

A significant challenge in assessing the stability of a stablecoin lies in the complexity of its underlying mechanisms. While stablecoin protocols are typically open and publicly available, their intricate structures often make it difficult for users and researchers to fully understand how they function and to predict their behavior under various market conditions.

Stablecoins can generally be categorized into three main types based on their collateralization models and operational mechanisms~\cite{Kahya2022}.
\textit{Fiat-backed stablecoins}, such as USDC and USDT, are backed by reserves of traditional currencies and are the most widely adopted category. 
\textit{Crypto-backed stablecoins} rely on cryptocurrencies as collateral, typically requiring over-collateralization to account for the volatility of their backing assets.
\textit{Algorithmic stablecoins} employ algorithmic and incentive mechanisms to manage their value. Unlike their collateral-backed counterparts, algorithmic stablecoins often operate under-collateralized, relying on dynamic adjustments to supply and demand rather than reserves of assets.

In this paper, we focus on \textbf{DAI}~\cite{team2020maker}, a stablecoin that straddles the line between crypto-backed and algorithmic models. Specifically, we present a formalized logical representation of DAI's mechanisms and policies, accompanied by a Prolog implementation. 
This implementation, which we make available as open-source software\footnote{The source code of our implementation is available at: \url{https://zenodo.org/records/15094256}}, provides a structured framework for analyzing DAI's operation and evaluating its robustness through simulations and theoretical inquiry.

Our contribution offers a novel approach to understanding the internal workings of DAI.
By formalizing its mechanisms, we aim to provide valuable insights into its stability and potential vulnerabilities, equipping researchers and practitioners with tools to better evaluate its performance even in stressed scenarios.
In this paper, we map the key DAI mechanisms into logic programming statements, which we implement using Prolog. 
The Prolog formalization of the DAI stablecoin includes two main conceptual blocks. 
The first defines operations for managing vaults, repay debt, and manage collaterals.
The latter manages essential system variables, including system surplus and debt, DAI savings and stability fees, and collateral, surplus and debt auctions.
The formalization allows researchers to assess system behavior.

\section{Related work}

Since their rise to prominence in 2018, stablecoins have been the subject of a growing body of literature aimed at predicting and understanding their inherent `stability' (e.g.,~\cite{bullmann2019search,arner2020stablecoins,ante2023systematic}).
The most prevalent approach has been data-driven and relies on the analysis of historical time series to assess current and future stability (e.g.,~\cite{kjaer2021empirical,duan2023instability,hoang2024stable}).
A second line of research has focused on constructing simulation models, typically either agent-based or equation-based, to replicate market behaviors and stress-test stability mechanisms of stablecoins (e.g.,~\cite{klages2020while,bhat2021daisim,kirillov2022stablesims,hafner2024thefour}). 
In contrast to these approaches, we propose to use a logic-based formulation to understand the complexity behind stablecoins' protocol with a particular focus on the DAI stablecoin.
To the best of our knowledge, no previous research has employed a logic-based framework for stablecoin analysis.

\section{Background}
DAI is a stablecoin, soft-pegged to USD, which is regulated and maintained by a decentralized autonomous organization: MakerDAO. Its peg is achieved through a series of market incentives and policies, and each actor involved in the process plays a key role in maintaining it. 

\begin{figure*}
    \centering
    \includegraphics[width=\columnwidth]{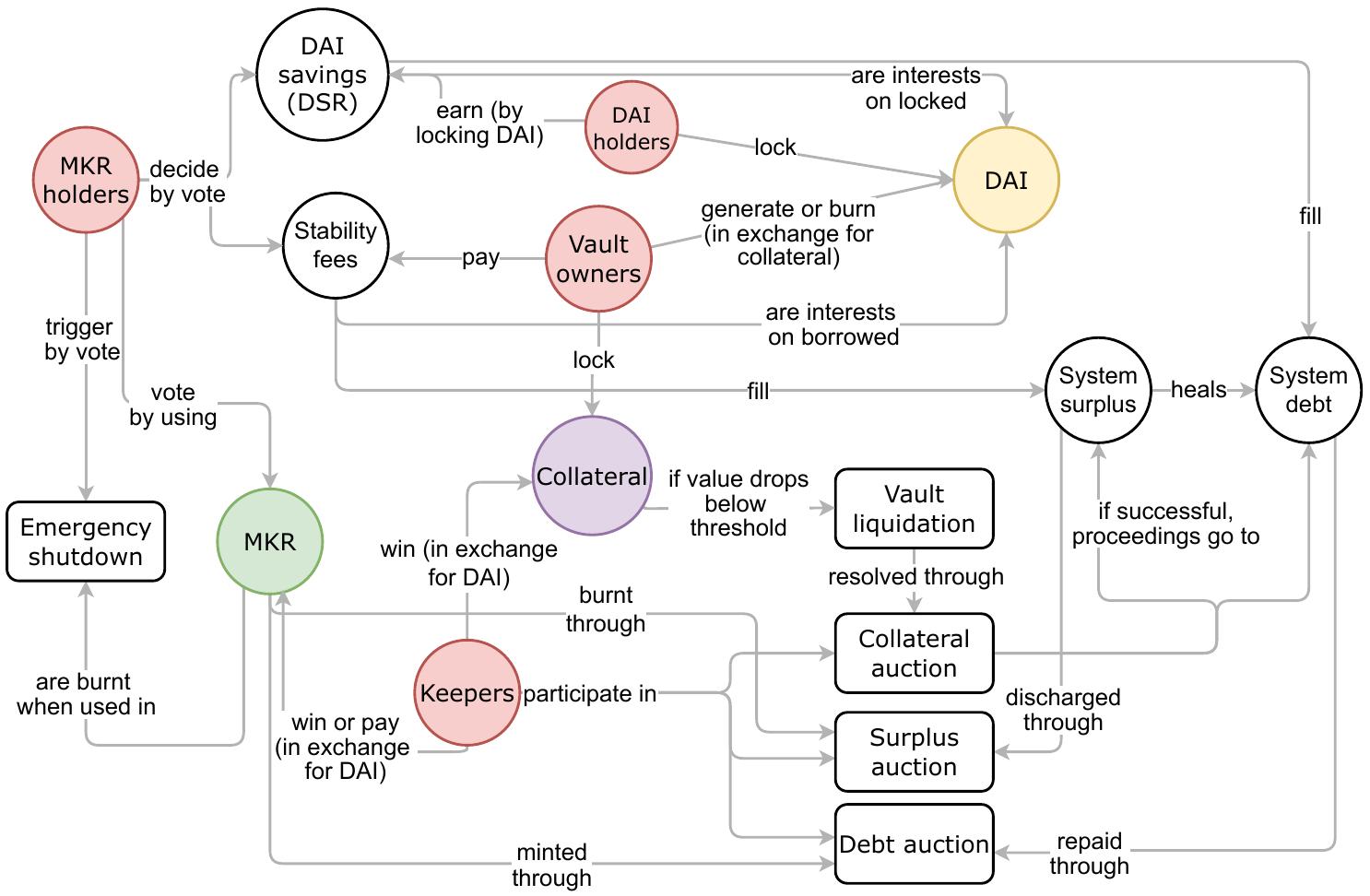}
    \caption{The overall picture of how the key concepts of DAI are interconnected. Red circles represent actors; colored circles are the three types of tokens (i.e., DAI, MKR, collateral); white circles are variables in the system; and rectangles denote actions or events}
    \label{fig:diagram}
    \Description[Key entities and their logic interconnection in DAI]{The overall picture of how the key concepts of DAI are interconnected.}
\end{figure*}

Figure~\ref{fig:diagram} summarizes the key concepts of DAI and the interactions among them.
The principal actors that contribute to the system are vault owners, DAI holders, MKR holders and keepers. The first three are internal actors, meaning they use core mechanisms of the system (like vaults), while the latter are external, because they are simply market agents.

\textbf{Vault owners} lock up assets in a vault as collateral to generate and borrow new DAI tokens, or return and destroy their DAI tokens to unlock their assets, thus varying the overall DAI supply in the system.
\textbf{DAI holders} own DAI and can lock it in a smart contract to earn interest. The offered interest rate (the DAI savings rate) can be changed through governance votes to influence DAI demand.
\textbf{MKR holders} own governance tokens (MKR) which they use to vote on the aforementioned interest rate, on vault parameters and on other decisions, to keep the system stable and handle risks and emergencies, thus increasing trust in the system.
\textbf{Keepers} are market actors (typically bots) who take advantage of arbitrage opportunities, making DAI price shift towards its target price, or trigger and participate in auctions to win collateral, DAI or MKR tokens, contributing in this way to minimize the amount of DAI debt in the system not backed by collateral.

\section{Issuance and Redemption of DAI}
The role of vaults is crucial to DAI: it is the sole mechanism through which DAI is issued and redeemed 
(as an analogy, this mimics the issuance of credit against collateral like in the case of mortgages, albeit in a decentralized and autonomous way).
In fact, whenever a vault owner borrows DAI, new DAI tokens are minted for them; conversely, whenever they pay their debt back, the paid DAI tokens are burnt.

To cover the risks inherent in this mechanism, borrowed DAI must be backed by collateral, which can be chosen among some tokens living on the Ethereum blockchain (ETH, wrapped BTC, etc.). This ensures that the borrowed DAI has a corresponding value in another token as a guarantee for insolvency. Moreover, vault owners must abide by some rules:
\begin{itemize}
    \item To reduce risks associated with fluctuating prices, their vaults must be overcollateralized; this means that the ratio between the value of the locked collateral and the amount of borrowed DAI (\textbf{collateralization ratio}) must be over a fixed threshold (\textbf{liquidation ratio}), which is strictly higher than 1; if a vault goes below such a threshold, it can be liquidated by a keeper and the underlying collateral can be put up for auction;
    \item They must pay interest (\textbf{stability fees}) on their loan, which serves as a risk premium and is mainly used for system cost coverage; they are free to pay interest when they pay their debt back, but cannot wait indefinitely because stability fees continuously accrue to their vault debt, thus lowering its collateralization ratio;
    \item They cannot borrow more DAI in a vault, if the DAI supply among all vault users for that vault type exceeds a fixed upper limit (\textbf{debt ceiling}); this keeps the overall supply for DAI in check; there is also a global debt ceiling which is a hard limit on all DAI regardless of the vault type.
\end{itemize}
%
Each collateral has several types of vaults associated with different parameters (usually trading higher stability fees for a lower liquidation ratio or vice versa). The parameters are set through governance votes based on risk considerations on the underlying collateral.

These conditions are exemplified by the following \texttt{vault\_create} predicate, which is true if the amount of borrowed DAI is within some fixed bounds (debt floor, debt ceilings) and its value against the collateral value is above the allowed threshold (liquidation ratio).

\begin{minted}{prolog}
vault_create(Vault_id, OwnerId, CollateralAmount, CollateralAsset, VaultType, IssuedCurrencyAmount) :-
  (var(Vault_id) -> gensym(vault, Vault_id); true),
  % check that a vault with same id doesn't exist
  \+vaults(Vault_id,_,_,_,_,_),
  % check that the collateral/vault type is a legal choice,
  vault_type(CollateralAsset, VaultType),
  % check debt ceilings and debt floor
  global_debt_ceiling(GlobalDebtCeiling),
  debt_ceiling(VaultType, DebtCeiling),
  debt_floor(VaultType, DebtFloor),
  debt_counter(VaultType, TotalDebt),
  global_debt_counter(TotalGlobalDebt),
  IssuedCurrencyAmount >= DebtFloor,
  NewTotalGlobalDebt is TotalGlobalDebt + IssuedCurrencyAmount,
  NewTotalDebt is TotalDebt + IssuedCurrencyAmount,
  NewTotalGlobalDebt =< GlobalDebtCeiling,
  NewTotalDebt =< DebtCeiling,
  % check collateralization ratio
  liquidation_ratio(VaultType, LiquidationRatio),
  collateral_price(CollateralAsset, Price),
  CollateralizationRatio is CollateralAmount * Price * 100 / IssuedCurrencyAmount,
  LiquidationRatio =< CollateralizationRatio,
  % create vault
  assertz_once(vaults(Vault_id, OwnerId, CollateralAmount, CollateralAsset, VaultType, IssuedCurrencyAmount)),
  % update debt counter
  set_debt_counter(VaultType, NewTotalDebt),
  set_global_debt_counter(NewTotalGlobalDebt).
\end{minted}

Similar predicates rule other operations on vaults: \texttt{vault\_withdraw\_collateral}, \texttt{vault\_generate\_dai} and \texttt{vault\_repay\_debt}, checking the same conditions only when necessary (for example, when withdrawing collateral, only the liquidation ratio must be checked).

Stability fees are calculated and accrued through the following:
\begin{minted}{prolog}
vault_add_stability_fees(Vault_id) :-
  vaults(Vault_id, OwnerId, VaultAmount, CollateralAsset, VaultType, IssuedCurrencyAmount),
  stability_fee_rate(VaultType, FeeRate),
  Interest is IssuedCurrencyAmount * FeeRate / 100,
  NewCurrencyAmount is IssuedCurrencyAmount + Interest,
  retractall(vaults(Vault_id, _, _, _, _, _)),
  assertz_once(vaults(Vault_id, OwnerId, VaultAmount, CollateralAsset, VaultType, NewCurrencyAmount)),
  % update debt counters
  debt_counter(VaultType, TotalDebt),
  global_debt_counter(TotalGlobalDebt),
  NewTotalDebt is TotalDebt + Interest,
  NewTotalGlobalDebt is TotalGlobalDebt + Interest,
  set_debt_counter(VaultType, NewTotalDebt),
  set_global_debt_counter(NewTotalGlobalDebt),
  % update vow balance adding fees as surplus
  vow_balance(Balance),
  NewBalance is Balance + Interest,
  update_vow_balance(NewBalance).
\end{minted}
This is meant to be used in conjunction with the previous predicates, which will return either true or false, if their conditions are met after accruing stability fees. The \textit{vow} balance is related to system surplus and debt and it will be explained later.

\subsection{Incentives for Vault Owners}
Since DAI works in a decentralized manner, each actor involved is independent and works for their own profit. This means that, in order for the system to function, each actor needs incentives to participate. Part of understanding DAI’s functioning is understanding these incentives.

Vault owners are incentivized to open a vault and borrow DAI mainly for three reasons:
\begin{itemize}
    \item They are looking to draw \textit{liquidity} from their assets and, since DAI is (promised to be) a stablecoin, it is a quick way to receive a loan in DAI and, through the DeFi environment, easily convert it in USD if necessary, without having to deal with intermediaries or strict payment schedules (they only need to take care of their collateralization ratio);
    \item They are trying to increase \textit{exposure} to their assets, by locking them in vaults and borrowing DAI against them, then using those DAI to buy more assets and repeating the process until it is no longer feasible (at some point the collateralization ratio will become inevitably close to the liquidation ratio, making the operation too risky); this strategy magnifies both potential losses and potential profits, the latter being the reason vaults attract potential investors;
    \item They want to earn \textit{income} by taking advantage of the difference (basically a \textit{yield spread}) between stability fees and the returns they can get by depositing their DAI in other DeFi platforms (like Yearn) with profitable interest;
    \item They want to take advantage of \textit{arbitrage} opportunities in the market.
\end{itemize}
Since vault owners are investors, they will react to DAI’s changing prices by varying the supply of DAI, thus making those prices shift back to the target price. In fact, whenever DAI’s price drops, they will buy more DAI to repay their debt at a discount, resulting in burnt DAI. Conversely, whenever DAI’s price rises, they will increase their debt\footnote{This is because the collateralization ratio is calculated using the \textit{amount} of generated DAI in that vault, not its current value in USD. This is done in order to make the market forces shift the price towards its target price of 1 USD.} to receive DAI to invest, resulting in newly minted DAI. Of course, this alone would make DAI be very volatile and would not be enough to maintain its peg in a predictable way. This is why other control mechanisms are needed and they will be discussed later.

\section{Control Mechanisms}
As we have seen, vaults are the source of DAI, and their owners contribute to keep the price at equilibrium by shifting its supply. This is heavily variable depending on the actual market conditions and that is why governance actions are needed to keep the price stable and reduce its oscillation. These actions change, through votation, important parameters involved in the process: the DAI Savings Rate and vault parameters. 

\subsection{DAI Savings Rate}
DAI holders can use a smart contract to deposit DAI into a container, called Pot, and earn savings based on a compound interest rate, called \textit{DAI Savings Rate (DSR)}. Users are free to deposit or withdraw DAI and collect their interest at any time, with no cost attached\footnote{However, since smart contracts require transactions on the Ethereum blockchain, each movement requires a fee (Ethereum Gas). Therefore, even if DAI holders may move their DAI freely from their Pot, there are costs inherent to these transactions, which discourage them from doing so too frequently.}. Interest earned by DAI holders is funded by stability fees paid by vault owners or, more generally, by the system surplus (which is discussed later).
Understanding incentives for DAI holders who make use of the Pot is straightforward: they save money and earn interest; this is similar to a time deposit but with no specific maturity date. This incentive is only for DAI holders who buy DAI and not for vault owners who take DAI from loans; in fact, there would be no profit for the latter (actually, there would be a loss), since vault stability fee rates must be structurally higher than the DSR, in order to cover interest paid to DAI holders (otherwise there would be unbacked debt in the system).
The DSR is a tool through which governance influences DAI demand; in fact, decreasing or increasing the DSR will translate, respectively, to a lower or higher demand of DAI.
%
The entire process can be compared to the \textit{monetary policy} implemented by central banks when they change interest rates. In this analogy, the stability fee on DAI vaults is akin to the base interest rate, while the savings rate on locked DAI is akin to a risk-free interest rate offered by government bond.

Earning interest through DSR is modeled by the following predicate:
\begin{minted}{prolog}
add_dai_savings(AddressId) :-
    dai_savings_rate(Rate),
    dai_deposit(AddressId, Amount),
    Interest is Amount * Rate / 100,
    NewAmount is Amount + Interest,
    update_dai_deposit(AddressId, NewAmount),
    % the vow's sin (unbacked debt) is increased, i.e. the vow balance is decreased by the previous interest
    vow_balance(Balance),
    NewBalance is Balance - Interest,
    update_vow_balance(NewBalance).
\end{minted}
Again, the vow balance will be explained later.

\subsection{Vault Parameters}
Vault parameters are another tool in the hands of MKR holders, since they affect the vault owners’ behavior and, thus, the supply of DAI. Since there are multiple vault types for each collateral,governance decisions can directly influence the supply of DAI coming from specific vault types, and not only the overall supply, allowing for better precision to react to risks arising from specific collaterals. For example, governance may want to induce a shift of the supply from a collateral type to another one, rather than a reduction or increase in the overall supply.

Let us now recall the main parameters:
\begin{itemize}
    \item \textbf{Stability fee rate} and \textbf{liquidation ratio} are risk parameters which determine how much interest a vault owner must pay and how much collateral they must lock in the vault; there may be different combinations of the two parameters on different vault types related to the same collateral, catering to different type of investors, who may be more or less prone to risks, allowing for a bigger exposure at a higher cost, or the opposite (for example, ETH-A type vaults might have 1\%/170\%, ETH-B might have 2\%/150\% and ETH-C might have 3\%/130\%); these values can be modified to incentivize generation or repayment of DAI, to encourage or discourage sudden influx to a particular collateral, or to trigger liquidations under exceptional situations;
    \item \textbf{Debt ceiling} puts a hard limit to the amount of DAI supplied by vaults of a specific type; this can be lowered to restrict influx of DAI coming from a single source, due to specific market conditions (usually following arbitrage opportunities), and can be raised again under safe conditions; 
    \textbf{global debt ceiling} exists to limit the overall supply of DAI; debt ceilings are also a last resort for when stability fee rates and liquidation ratios cannot be modified (for example, if they are too low);
    \item \textbf{Liquidation penalty} is a fee which vault owners must pay if their vault is liquidated (as a consequence of going below the liquidation ratio); it is paid indirectly by cutting a bigger portion of the collateral during a collateral auction; it contributes to stability because it encourages vault owners to keep their vaults overcollateralized, by introducing an additional penalty for liquidations, and it’s used to cover the rewards given to keepers for triggering liquidations\footnote{Keepers are incentivized to trigger liquidations for undercollateralized vaults by receiving a fee (as a linear function of the vault size). Liquidation penalties cover these costs and also avoid that a malicious vault owner exploits the system by triggering liquidations and winning auctions on their own vault.}.
\end{itemize}

Vault parameters can be modeled by dynamic predicates which are used to look up the correct value for the parameter. For example, the following declares the stability fee predicate:

\begin{minted}{prolog}
    :- dynamic stability_fee_rate/2.
\end{minted}

We have already seen these predicates being used in the following way inside other predicates, to retrieve the value:

\begin{minted}{prolog}
    stability_fee_rate(VaultType, FeeRate)
\end{minted}

The \texttt{governance} module is made of predicates which can dynamically assert the previous parameters (i.e. change their value). Using the stability fee rate example again:

\begin{minted}{prolog}
set_stability_fee_rate(VaultType, FeeRate) :-
    retractall(stability_fee_rate(VaultType, _)),
    assertz_once(stability_fee_rate(VaultType, FeeRate)).    
\end{minted}

The same is done for all other parameters, through dedicated dynamic predicates: \texttt{liquidation\_ratio}, \texttt{liquidation\_penalty}, \texttt{debt\_floor}, \texttt{debt\_ceiling}, \texttt{global\_debt\_ceiling} and \texttt{dai\_savings\_rate}.


\subsection{Other Parameters}
There are other parameters in the system which can be modified by governance votes, and they are briefly mentioned here for the sake of completeness. 
The \textbf{debt floor} is the minimum debt allowed for a vault; that is, a user cannot borrow less than that amount when opening a vault, and cannot reduce their debt below that amount when they are repaying a portion of their debt (unless they’re repaying the whole debt). Its purpose is to limit the generation of too many vaults with a low debt which may be inefficient to liquidate due to unprofitable conditions (for example, because of transaction fees, i.e. Ethereum Gas).
Other parameters are concerned with other aspects of the system: some regulate bids and durations in \textbf{auctions}, while others bound the size of the \textbf{system buffers} (pools for system surplus and system debt, which are explained later).

\section{Crisis Resolution Mechanisms}

\subsection{Liquidating a Vault}
Under normal conditions, everything is meant to avoid liquidations. In fact, if no liquidations are needed, it means that all vaults are overcollateralized, keeping the system stable. That is why keeping the right incentives through governance decisions is of primary importance. However, market unpredictability can rapidly change the price of collaterals underlying DAI vaults. If it shifts too much, the collateralization ratio of a vault might drop below the liquidation ratio, signaling that there’s a risk for unbacked debt. Since these situations are usually temporary, governance voters don’t respond to them. Even under the necessity of governance action, the voting process and change implementation is not immediate, as we will see later. Therefore, governance decisions can change incentives and shift equilibrium in the long run, but cannot react quickly to market fluctuations or sudden crises. That is why a mechanism to cover possible unbacked debt is needed. This is done through liquidations and, consequently, \textbf{collateral auctions}.

When a vault accrues too much debt (due to a risky strategy or market conditions), its collateralization ratio drops below the allowed liquidation ratio for that type of vault. This does not mean that there is unbacked debt, because the liquidation ratio is always strictly greater than 1, but to avoid ever reaching a point where there are more DAI than its corresponding value in collateral, the MakerDAO community aims at keeping the overall system overcollateralized. This means that, individually, all vaults must have more value in collateral than they borrowed in DAI. Therefore, if they are below their liquidation ratio, they are marked as risky and anyone can \textit{bite}\footnote{This is the technical term in the docs when referring to triggering liquidations.} them to initiate a liquidation. This is done by market actors, called keepers.

When a vault is liquidated, it is closed and its underlying collateral is put up for auction. Keepers participate in this collateral auction to win collateral at a profitable price and, in turn, they pay in DAI to cover the collateral value plus a liquidation penalty. Vault owners only get back the remaining collateral, if any. In this way, liquidations act both as a deterrent (so that vault owners keep their vault in check) and as a resolution mechanism in case of need.

Here is an exemplification of the process:
\begin{minted}{prolog}
liquidate_vault(Vault_id) :-
    % check if vault can be liquidated
    liquidation_condition(Vault_id),
    vaults(Vault_id, _, CollateralAmount, CollateralAsset, VaultType, Debt),
    % liquidation is initiated: debt is removed and collateral put on auction
    vault_repay_debt(Vault_id, Debt),
    vault_withdraw_collateral(Vault_id, CollateralAmount),
    liquidation_penalty(VaultType, Rate),
    TotalDebt is Debt * (1 + Rate / 100),
    % if auction is successful, remaining collateral is given back to the vault owner and proceedings are put into the vow surplus
    % else all collateral is given back to the the vault owner but its debt becomes "bad debt" which increases the vow debt (sin)
    vow_balance(VowBalance),
    (
        collateral_auction(CollateralAmount, CollateralAsset, TotalDebt, RemainingCollateral, Proceedings) -> 
        (
            vault_deposit_collateral(Vault_id, RemainingCollateral),
            IncreasedBalance is VowBalance + Proceedings,
            update_vow_balance(IncreasedBalance)
        );
        (
            vault_deposit_collateral(Vault_id, CollateralAmount),
            DecreasedBalance is VowBalance - TotalDebt,            update_vow_balance(DecreasedBalance)
        )
    ).    
\end{minted}

The vow balance is explained in the next section, while the collateral auction is modeled by the following:
\begin{minted}{prolog}
collateral_auction(CollateralAmount, CollateralAsset, TotalDebt, RemainingCollateral, Proceedings) :-
    direct_auction(CollateralAmount, DaiOffered, CollateralAsset, "DAI"),
    DaiOffered >= TotalDebt,
    reverse_auction(CollateralReceived, DaiOffered, CollateralAsset, "DAI"),
    CollateralReceived > 0,
    RemainingCollateral is CollateralAmount - CollateralReceived,
    Proceedings is DaiOffered - TotalDebt.    
\end{minted}

The \texttt{direct\_auction} and \texttt{reverse\_auction} predicates are abstractions of how two different models of auction work, and must be implemented with specific rules. More details on auctions are found in the appendix.

There are a few important implications in the payment made by the keepers:
\begin{itemize}
    \item They are basically paying back the vault debt in place of the vault owners; this means that, when they do, that DAI is burn
    \item It may look like the liquidation penalty weighs on the keepers, which would make the auction less appealing; however, keepers bid on how much collateral to accept for a price equal to the debt plus the penalty and, thus, the presence of the penalty will decrease the amount of collateral remaining after the auction; this means that vault owners indirectly pay their penalties in the end
\end{itemize}

\subsection{System Debt and System Surplus}
As we have seen, liquidations and collateral auctions are a tool used to avoid accumulating too much unbacked debt in the system. However, even collateral auctions can fail. In fact, if market conditions are not profitable for keepers, no bid will ever cover the outstanding debt plus the liquidation penalty. In this case, the vault owner’s debt is still unpaid and is not backed up by collateral. This debt goes to fill a system debt buffer, so that it can be tracked and later recovered. This mechanism has been expressed in the previous section by the \texttt{liquidate\_vault} predicate.

More generally, \textbf{system debt} consists of all “bad debt” (called \textit{sin}), i.e. debt that is not backed by any collateral. This can be interest paid to DAI holders or debt coming from failed collateral auctions.
Analogously, the system keeps track of the surplus in another buffer. \textbf{System surplus} consists of stability fees and liquidation proceedings (which also contain liquidation penalties).
These two pools cancel each other to determine whether the system has net debt or net surplus. The fact that stability fees cover interest paid on DAI savings is a particular case of this mechanism.

The “balance sheet” tracking debt and surplus is the vow contract, which we have modeled through a dynamic predicate:

\begin{minted}{prolog}
% internal balance sheet (Vow = Surplus - Sin)
:- dynamic vow_balance/1.  
\end{minted}

As we have seen, this is updated when adding stability fees, earning DSR, by liquidations or by auction resolutions, through the following assertion:

\begin{minted}{prolog}
 update_vow_balance(Balance) :-
    retractall(vow_balance(_)),
    assertz_once(vow_balance(Balance)).   
\end{minted}

When net debt or net surplus is over a threshold (chosen by governance), it must be auctioned, in exchange for governance tokens (MKR). In the first case, new MKR are minted in exchange for DAI to cover the debt (\textbf{debt auction}). In the second case, surplus DAI is auctioned off in exchange for MKR which are burnt (\textbf{surplus auction}). This can be expressed in the following way:

\begin{minted}{prolog}
 debt_auction(DaiToPay) :-
    get_net_debt(Debt),
    Debt >= DaiToPay,
    reverse_auction(MkrReceived, DaiToPay, "MKR", "DAI"),
    % mint new MKR
    mint_maker(MkrReceived),
    % update vow balance by removing paid debt
    vow_balance(Balance),
    NewBalance is Balance + DaiToPay,
    update_vow_balance(NewBalance).   
\end{minted}

\begin{minted}{prolog}
surplus_auction(DaiAuctioned) :-
    get_net_surplus(Surplus),
    Surplus >= DaiAuctioned,
    direct_auction(DaiAuctioned, MkrOffered, "DAI", "MKR"),
    % burn paid MKR
    burn_maker(MkrOffered),
    % update vow balance by removing surplus
    vow_balance(Balance),
    NewBalance is Balance - DaiAuctioned,
    update_vow_balance(NewBalance).
\end{minted}

It is worth mentioning that, while there is a certain symmetry between debt and surplus, having system debt is much more risky than having surplus. For this reason, the debt buffer is much smaller than the surplus buffer, shifting the equilibrium in favor of surplus. This means that, under normal circumstances, the system will always have a pool of surplus through which it heals the debt, and debt auctions are left as a last resort mechanism.

Even when the decisions of market actors and governance shift DAI’s price towards its target price, there are still exceptional situations which are not covered: irrational market, critical bugs, security breaches or, more simply, necessary protocol upgrades all require that the governance act immediately. In these cases, an \textbf{emergency shutdown} is initiated. For example, this can be useful in case of unexpected price drops of one or more collaterals, price errors or attacks to Oracle price feeds (more on Price Oracles in the Appendix), Governance attacks (like malicious proposals), loss of trust in the system which results in panic selling, and so on. In these situations, traditional voting is too slow to react effectively.

MKR voters can trigger the emergency shutdown, by locking their MKR into a designated smart contract (the Emergency Shutdown Module, or ESM). If there is a quorum of votes, the procedure is started and the MKR tokens are burnt. This is modeled by the following predicates:

\begin{minted}{prolog}
% The process of initiating Emergency Shutdown is decentralized and controlled by MKR voters, 
% who can trigger it by depositing MKR into the Emergency Shutdown Module.

majority(Account):-
  shares_total_issuance(Total_issuance),
  shareholder_accounts(Account, Balance),
  Balance > Total_issuance/2.

esm_active :- 
  majority("esm").
\end{minted}

The shutdown procedure then works as follows:
\begin{enumerate}
    \item The system is frozen: vault operations are disabled, debt and surplus auctions are disabled, payment of interest on DAI savings is suspended (i.e. DSR is set to 0), prices are frozen to their current value, votes are only allowed if related to the Emergency Shutdown process or to the future redeployment of the system;
    \item Owed debt from vaults is canceled and the underlying collateral is taken; vault owners can claim the remaining excess collateral;
    \item A cooldown period is set to handle undercollateralized vaults; within this timespan liquidations can still be triggered and ongoing collateral auctions can be concluded;
    \item Final collateral price is adjusted considering system surplus/debt, and DAI tokens can be exchanged for a relative share of all types of collaterals present in the system; this exchange is made using DAI \textbf{target price} of 1 USD.
\end{enumerate}

ESM effect on vaults can be expressed thus:

\begin{minted}{prolog}
esm_effect :-
  foreach(
    vaults(Vault_id, _, _, _, _, _),
    vault_withdraw_excess_collateral(Vault_id)
  ).
\end{minted}

where the bond with the target price is made explicit in the following predicate:

\begin{minted}{prolog}
vault_excess_collateral(Vault_id, Amount, Asset) :-
  vaults(Vault_id, _, C_amount, Asset, _, I_amount),
  collateral_price(Asset, C_price),
  target_price(T_price),
  C_needed_to_refund_currency is I_amount*T_price/C_price,
  Amount is C_amount-C_needed_to_refund_currency.
\end{minted}

The shutdown procedure blocks any action that could create more debt in the system, freezes the prices at the time of the shutdown and gives everyone (vault owners and DAI holders) the collateral they are entitled to. This expectation is finally what drives DAI’s price to its target price: even when everything goes wrong, 1 DAI token can be used to redeem 1 USD worth of collaterals.
However, there’s a caveat: since there is no guarantee that MKR tokens will hold their value after shutdown, there’s no possibility to discharge any surplus or debt through standard auctions. That is why surplus and debt auctions are disabled and any surplus or debt is passed on to DAI holders, who will receive it as a reward or penalty when they redeem the collateral (this is reflected by the adjusted collateral exchange price, mentioned in the procedure above).
It is also worth mentioning that vault owners are prioritized over DAI holders, since they can withdraw their excess collateral before DAI can be exchanged. This is done because the exchange ultimately is affected by the overall collateralization of the system, so each vault owner needs to retrieve their excess collateral based on their collateralization; otherwise, they would be less incentivized to keep vaults overcollateralized or to pay stability fees.

\section{Governance}

\subsection{Role and Incentive}
In the previous sections, we have seen how the system reacts to market fluctuations to keep its peg, in the long run (changing parameters), in the short run to absorb bad debt (auctions), and during emergencies (shutdown). Governance has a key role in these control mechanisms and in making sure that the system is safe and keeps working. Let us summarize their main tasks here:
\begin{itemize}
    \item Decide which collaterals are allowed and decide vault parameters by vault type for each collateral;
    \item Modify the DAI Savings Rate;
    \item Change auction parameters;
    \item Handle emergencies through shutdown;
    \item Choose price feeds;
    \item Update the system with protocol changes, security upgrades, etc.
\end{itemize}

Therefore, the governance acts as policy makers to keep the system stable. Their decision process is decentralized through a voting system. Ultimately, this is what makes the system decentralized, since there are no centralized policies, because anyone can participate in voting as long as they hold MKR tokens. Essentially, each token is one vote (the voting mechanism is explained more in detail in the Appendix). This also reveals an incentive for MKR holders: keeping high the trust in the system, which inevitably reflects on the market value of the very tokens they hold.

\subsection{MKR Tokens}
Supply of MKR tokens is regulated by auctions and, in our model, it is a dynamic predicate, modified by auctions through \texttt{burn\_maker} and \texttt{mint\_maker} predicates.

\begin{minted}{prolog}
% total maker supply
:- dynamic maker_supply/1.
\end{minted}

While a fixed amount of MKR was distributed when the system launched, the only way to issue (mint) and redeem (burn) MKR now is through debt auctions (by paying DAI) and surplus auctions (by receiving DAI) respectively.
This suggests another goal for MKR tokens, in addition to voting: they are an incentive for keepers to pay off bad debt present in the system and, in turn, get new MKR as a reward. On the other hand, this encourages MKR holders to govern responsibly to avoid too much debt in the system which would cause an increase in MKR supply.

In conclusion, MKR tokens serve two purposes: governance and recapitalization. However, this keeps working as long as MKR tokens have value. Since they are influenced by the value of DAI and vice versa (through auctions), if they were the only two types of tokens involved in the process, it could cause a death spiral if one of the two dropped considerably. This is why a third element is necessary in the equation to keep the system working: collaterals, which leads us back to where we started in this discussion.

\section{Example}
We briefly illustrate how the proposed framework can be used to simulate the functioning of DAI.
We use \mintinline{Prolog}{initialize_system} to define (default) system parameters (e.g., prices, exchanges rates, type of vaults). It creates vault types (including ETH-A and ETH-B) and set  ETH's value to $150$~USD. ETH-A has liquidation ratio $150\%$,  ETH-B $130\%$.

With the following snippet, we simulate a user (ID $200$) that creates a vault (ID $1$) of type ETH-A by depositing $2$~ETH and borrowing $150$~DAI (this is allowed by the ETH price in USD and the liquidation ratio set in our example). 
%
Then, we simulate a sudden price drop (ETH to $45$~USD), which makes the vault liquidate.  
The price drops so low that the collateral auction is not profitable and the debt goes to fill the balance. As expected, \mintinline{Prolog}{vow_balance} is now negative.
\begin{minted}{prolog}
initialize_system(),
% vault 1, User 200 puts 2ETH and creates 100DAI
vault_create(1,200,2,"ETH","ETH-A",100),
% sudden price drop
collateral_set_exrate_and_price("ETH",45),
liquidate_vault(1),
vow_balance(V).
\end{minted}

Consider now that, due to the big price drop, another user (ID $201$) buys a lot of ETH and later, when price is back to normal, uses it to borrow $2300$~DAIs (this is feasible for the $130\%$ liquidation ratio of ETH-B).
The user played it risky and, when stability fees kick in, the vault goes just below the liquidation ratio. This time the price is high enough for a collateral auction to be profitable, and we can check that the balance goes back to positive, due to the auction proceedings.

\begin{minted}{prolog}
% price back to normal
collateral_set_exrate_and_price("ETH",150),
% vault 2, User 201 puts 20ETH and borrows 2300DAI
vault_create(2,201,20,"ETH","ETH-B",2300),
% sudden price drop
collateral_set_exrate_and_price("ETH",45),
% stability fees are added to debt
vault_add_stability_fees(2),
liquidate_vault(2),
vow_balance(V).
\end{minted}
%

Since both these scripts return \mintinline{Prolog}{true}, we are sure that all the actions were legal, and we could add new checks like \mintinline{Prolog}{V > 0.}, without worrying about the parameter constraints (e.g., vault can be liquidated, collateral auction is feasible).

\section{Conclusion}
The previous discussion shows that the way DAI keeps its peg is through the combined action of external market forces and internal incentives, modified through governance action. Moreover, the system has two layers of defense against bad debt (collateral actions and debt actions) and an emergency resolution mechanism. All these mechanisms contribute to the stability of the system in the long run or in case of immediate crisis, and their functioning depends on the existence of the collateral and the MKR token. While the first one covers debt, the second one is key to the voting system and covers debt in exceptional situations. But even during emergencies, MKR could be worth nothing and what keeps DAI’s price from falling is the promise that it can be exchanged for an equivalent value of collateral. All considered, the collateral is the determining factor in maintaining the peg and all mechanisms ensure that for every DAI there is some corresponding collateral in the system. This means that the peg is directly dependent on the value of the underlying collaterals, meaning that, if a collateral goes down, it could bring DAI along. The only way to minimize this effect is to diversify the collateral distribution as much as possible, which has been done by introducing different collateral types (including even real-world assets).
Trust in the system is another important factor to consider for MKR to have any value at all. The transparency and decentralization of the voting system, together with extensive documentation to help people make informed decisions, is surely a good step in that direction.
Lastly, while the system still keeps some points of centralization, like the Maker Foundation, their actions are limited and most decisions are done in a decentralized way, thanks to how the decision process is handled.

In this paper, we have presented a formalization of DAI's core mechanisms using Prolog logic programming. Our framework models the protocol's fundamental operations, enabling users to simulate actions and validate their effects on the system’s state. By providing this formalization, we offer a tool for exploring DAI’s behavior and identifying potential vulnerabilities. Future work can leverage this framework to conduct more detailed analyses of the system under various scenarios, contributing to a deeper understanding of its long-term stability and resilience.

\section*{Acknowledgements}
We would like to thank Luigi Bellomarini for giving us the opportunity to investigate this topic and suggesting to use a logic-based approach to investigate the mechanisms of a stablecoin. 
We would also like to extend our thanks to Stefano Sferrazza for his initial contributions and insights.

\bibliographystyle{ACM-Reference-Format}
\bibliography{biblio}

\newpage
\appendix

\section{Supplemental Material on Relevant Mechanism and Further Actors}

\subsection{Auctions}
There are three types of auctions in the system: collateral auctions, debt auctions and surplus auctions.

\textbf{Collateral auctions} are triggered after vault liquidations. When a vault is bitten, all its collateral is put up for auction (or a portion of it, if it’s more than the allocated lot size for the auction). Then, the auction proceeds in two phases:
\begin{enumerate}
    \item  \textit{Direct auction}: the participants bid increasing amounts of DAI to buy the auctioned collateral; if there is a bid that covers the outstanding debt plus the liquidation penalty, the auction proceeds to the next phase; otherwise, the system debt increases
    \item A \textit{reverse auction}: the bid decided in the previous phase is now fixed and the participants bid on decreasing amounts of collateral they would buy for that fixed price; this is done to minimize losses for vault owners
\end{enumerate}
When the auction is resolved, the collateral is sent to the winner, the paid DAI is burnt and any residual collateral is sent to the vault owner.

\textbf{Debt auctions} and \textbf{surplus auctions} are triggered under similar but opposite circumstances. Whenever a user sends a \textit{heal} transaction, system debt is netted out from system surplus. If net debt or net surplus is larger than a given threshold, the exceeding amount of DAI is put up for auction in blocks of fixed size (\textit{lot}). Then, the auction proceeds in different ways:

\begin{itemize}
    \item For debt auctions, the fixed amount of DAI is to be paid, and the participants bid on decreasing amounts of MKR they are willing to accept in exchange for that DAI (a \textit{reverse auction}); paid DAI is then burnt and won MKR is newly minted
    \item For surplus auctions, the fixed amount of DAI is put up for sale, and the participants bid with increasing amounts of MKR to win it (a \textit{direct auction}); paid MKR is then burnt and DAI is sent to the winner
\end{itemize}

All auctions have some risk parameters, which were hinted at in a previous section: minimum bid increase (as a percentage), bid duration and auction duration. These parameters are decided by governance to make auctions efficient, i.e. resolve them successfully without dragging forever.

\subsection{Voting}
Proposals and votes happen in two ways: off-chain and on-chain. Off-chain voting happens on the MakerDao forum, where everyone can participate to make and discuss new proposals or talk about existing problems. However, final decisions are made on-chain using MKR tokens to vote. This is done in two steps: first, governance polls are made to evaluate popular proposals, usually already well-received on forums; then, an executive vote is cast to approve the changes favored in the polls and make them effective.

In both governance polls and executive votes, each MKR corresponds to one vote. Moreover, since votes are on-chain transactions, they require Ethereum Gas (i.e. a fee paid in ETH).
However, there is a key difference in how the two types of vote work: while MKR used to vote can be withdrawn in both governance polls and executive votes, it has no effect in the first case, but it does in the second one. The reason for this is that governance polls are used only to make decisions, not to make them effective and, thus, once they are closed, the final result cannot be changed and MKR can be released. On the other hand, executive votes are continuously subject to effective changes, because, whenever a new proposal conflicts with a previous one, it must compete against all existing proposals in the number of votes. This means that releasing MKR from a vote, even after changes have been made effective, weakens that decision, because a future executive vote could revert the changes more easily. For this reason, MKR used for executive votes are usually kept in the voting contract and their holders move them, if necessary, to new proposals (however, MKR holders are free to withdraw their MKR whenever they want, and they could also add new MKR for new votes). This behavior is also encouraged by the fact that, before casting an executive vote, MKR must be deposited into the voting contract, requiring another on-chain transaction (and, consequently, another Gas fee), while keeping MKR in the contract removes this necessity.

Votes can decide on system parameters, collaterals or oracles allowed, or they can change how the system itself works (for example, make changes to the protocol or the governance process). In the latter case, a Maker Improvement Proposal (MIP) must be voted and a quorum of voters is required, to avoid potential attackers passing arbitrary changes. The quorum threshold can be changed by governance with another MIP.

Once executive votes are closed, there is a time delay before changes are made effective. At the time of writing, this is 48 hours, but it can be changed by voting. This is handled by the Governance Security Module and its purpose is to thwart potential governance attacks, like malicious proposals.
Moreover, changes are public and can be triggered by anyone, after the executive vote has passed. However, a simple description in the executive vote would not be sufficient to guarantee that the implied changes would be corresponding to the description. That is why executive votes contain a hash of the contract which will be executed (this is called a \textit{spell}). This means that anyone can check the lines of code that will be executed if the vote passes, and anyone can trigger the \textit{spell} once the voting phase is concluded. Therefore, this mechanism helps in decentralizing the whole governance process.

\subsection{Oracles}
Pricing oracles are essentially external price feeds that allow to determine exchange rates needed for vault operations, auctions or emergency shutdowns. There is an Oracle module for each collateral, which takes market values from multiple sources and calculates a median. This is exemplified by the following:

\begin{minted}{prolog}
% collateral_price(asset, price)
% price is expressed in USD
:- dynamic collateral_price/2.

collateral_set_price(Asset, Price) :-
  retractall(collateral_price(Asset, _)),
  assertz_once(collateral_price(Asset, Price)).
\end{minted}

To prevent oracle attacks (like price manipulation), there are a few security mechanisms:
\begin{itemize}
    \item The sources from which the values are taken must be whitelisted;
    \item The final price is a median of the collection of prices: this allows the result to be more robust to outliers and, thus, pricing errors (whether intentional or not);
    \item An Oracle Security Module is implemented for each collateral, and it introduces a delay before the median price can be taken by the system.
\end{itemize}

\subsection{DAO Teams}
Beyond keepers and oracles, there are other external actors, who contribute to maintain the system active and functional. This group consists of individuals or organizations who are engaged (by governance through voting) to provide services to the platform: for example, the Risk Team evaluates financial risks, onboarding or offloading of collaterals, while Governance Facilitators support the governance process and communication. These are collectively known as DAO Teams.

\end{document}